\documentclass[journal]{IEEEtran}
\usepackage{amsmath,amsfonts}
\usepackage{amssymb}
\usepackage{algorithmic}
\usepackage{algorithm}
\usepackage[caption=false,font=normalsize,labelfont=sf,textfont=sf]{subfig}
\usepackage{textcomp}
\usepackage{stfloats}
\usepackage{url}
\usepackage{verbatim}
\usepackage{graphicx}
\usepackage{cite}
\usepackage{xcolor}
\usepackage{booktabs}
\usepackage{array}

\usepackage{acronym}
\newacro{gfm}[GFM]{Grid-Forming}
\newacro{gfl}[GFL]{Grid-Following}
\newacro{ibr}[IBR]{Inverter-Based Resource}
\newacro{pll}[PLL]{Phase-Locked Loop}
\newacro{vsm}[VSM]{Virtual Synchronous Machine}
\newacro{sm}[SM]{Synchronous Machine}
\newacro{cf}[CF]{Complex Frequency}
\newacro{coi}[CoI]{Center of Inertia}
\newacro{rocof}[RoCoF]{Rate of Change of Frequency}
\newacro{dfig}[DFIG]{Doubly-Fed Induction Generator}
\newacro{gvr}[GVR]{Global Voltage Regulation}
\newacro{qss}[QSS]{Quasi-Steady State}
\newacro{pmu}[PMU]{Phasor Measurement Unit}
\newacro{dc}[DC]{Data Center}

\begin{document}

\title{System-wide Dynamic Performance Metric for Converter-Dominated Power Systems}

\author{
    \IEEEauthorblockN{
        Rodrigo Bernal\IEEEauthorrefmark{1}, 
        Taulant K\"{e}r{\c{c}}i\IEEEauthorrefmark{2}, and 
        Federico Milano\IEEEauthorrefmark{1} \\
    }
    \IEEEauthorblockA{
        \IEEEauthorrefmark{1}School of Electrical and Electronic Engineering, University College Dublin, Belfield, Ireland \\
    }
    \IEEEauthorblockA{
        \IEEEauthorrefmark{2}EirGrid (Irish Transmission System Operator), Dublin, Ireland \\
    }
    \thanks{Corresponding author's e-mail: federico.milano@ucd.ie}
    \thanks{This work was partially supported by the Sustainable Energy Authority of Ireland (SEAI) by funding R.  Bernal and F.  Milano through the FRESLIPS project, Grant No.  RDD/00681.}
    \vspace{-3mm}
}




\maketitle
\begin{abstract}
  Fast controller dynamics in \acp{ibr}-dominated power networks couple frequency and voltage, making traditional decoupled analysis inadequate. We introduce a unified system-wide metric—the complex frequency of system losses—to evaluate dynamic performance and synchronization.  Composed of device- and network-driven terms, it expresses voltage and injected-current phasor variations weighted by bus complex powers. Validated on an \ac{ibr}-modified IEEE 39-bus system and the Irish transmission grid, the framework effectively assesses $R/X$ sensitivities, dynamic support limits, and data-center load impacts.
  %
\end{abstract}


\section{Introduction}

Voltage in AC power systems is commonly described as a phasor defined by two fundamental components: magnitude and frequency. Historically, these components have been assessed through different paradigms. Frequency has traditionally been treated as a global quantity, often evaluated using aggregated metrics like the frequency of the \ac{coi} to gauge overall dynamic performance \cite{Paganini2017, TAN2022, DULAL2026, Kou2014, Zhao2019, Milano2018, FERNANDEZ2019}. 
Nevertheless, recent research redefines frequency as an inherently local and instantaneous variable \cite{Milano2022cmplx}—a critical distinction in modern grids. The integration of \acp{ibr} on the generation side and converter-based demand (such as large-scale data centers) on the load side results in system dynamics that are substantially faster and more heterogeneous \cite{kerci2026local}.   

Accurate calculation of the \ac{coi} frequency relies on techniques for identifying power system inertia, categorized broadly into model-based and measurement-based approaches \cite{TAN2022,DULAL2026}. Model-based methods depend on dynamic models and parameter estimation, whereas measurement-based techniques range from large-disturbance approaches (e.g., estimating inertia from transient \ac{rocof} and power imbalances) to ambient-data methods (e.g., extracting inertia from continuous small fluctuations) \cite{Paganini2017,Kou2014}. Specific implementations include robust Kalman filtering for online \ac{coi} estimation \cite{Zhao2019} and direct algebraic calculations of the \ac{coi} frequency as a weighted sum of bus frequencies \cite{Milano2018}.  These approaches supply system operators with aggregated indicators—such as frequency nadir, \ac{rocof}, and inter-area oscillations—to evaluate overall dynamic behavior.

On the other hand, voltage magnitude has consistently been studied as a strictly local quantity. However, fast controllers operating within both \acp{ibr} and converter-based loads cause the time scales of voltage magnitude and frequency dynamics to overlap. This overlap raises the question of whether a system-wide, aggregated metric encompassing both quantities is necessary.  While frequency effectively assesses power imbalances, existing literature has expanded global variables to coordinate objectives beyond basic frequency regulation. Examples include global voltage quality metrics for distribution networks \cite{SAIED2001427}, unified nonlinear controllers for transient stability and voltage regulation \cite{Guo2001, Gordon2008}, consensus-based regulation of voltage drift in dc microgrids \cite{ZHAO201518}, and dynamic distributed slack buses formulated to handle system-wide power imbalances \cite{dynslack}.

Despite these advances, a unified system-wide metric that simultaneously captures voltage magnitude and frequency dynamics—particularly those driven by uncoordinated converter-based demand—remains an open gap. Conventional aggregated metrics consider frequency or voltage in isolation, failing to account for how voltage perturbations propagate through the network or how local voltage behavior interacts with high-variability power injections.

To fill this gap, we introduce a unified system-wide metric based on the complex frequency of total system losses. This formulation leverages total system losses to naturally aggregate voltage deviations, current flows, and network topology, inherently coupling voltage and frequency dynamics.  

The main contributions of this paper are as follows:
\begin{itemize}
\item We propose a unified metric based on the complex frequency of the total system losses to capture coupled voltage and frequency dynamics in networks driven by \acp{ibr} and converter-based demand.
\item We derive the metric from first principles, decomposing it into device- and network-level contributions to provide physical insight into dynamic drivers.
\end{itemize}


Analogous to traditional \ac{coi} frequency, the metric serves as an offline assessment tool, specifically aimed at evaluating the ability of the grid to retain system-wide synchronism in both normal and abnormal operating conditions. 

The framework is validated through two case studies: an \ac{ibr}-modified IEEE 39-bus system to evaluate sensitivity to the $R/X$ ratio and limited dynamic support, and the all-island Irish transmission system to assess grid stability under fast, uncoordinated data-center load variations \cite{kercci2024impact}.

\section{Nomenclature}
Unless otherwise stated, all variables are time-dependent.    Complex variables are denoted by $\bar{x}$.  Symbols $\, \dot{x} \,$ and $\bar{x}^{*}$ denote the time derivative and the complex conjugate of a variable, respectively.  For a complex variable in polar form $\bar{x}=x e^{\jmath\theta}$, its \ac{cf} is defined as $\bar{\eta}_x = \varrho_x + \jmath\omega_x$ with $\varrho_x = \dot{x}/x$ and $\omega_x = \dot{\theta}$, satisfying $\dot{\bar{x}} = \bar{x}\bar{\eta}_x$.

\section{Proposed System-wide Dynamic Metric}
Let the power injection at bus $h$ of an ac power system be:
\begin{align}
    \bar{s}_{h}=p_{h}+\jmath \, q_h, \quad \forall h \in [1,\,\dots\,,n],
    \label{eq_complex_power}
\end{align}
where $p_{h}$ and $q_h$ indicate the active and reactive powers injected at bus $h$, respectively, and $n$ is the number of network buses.  Assuming that loads are modeled as negative power injections, the sum of the powers injected across all buses describes the complex power losses of the system:
\begin{align}
    \bar{s}_{\mathrm{l}}  = \sum\nolimits_{h}^{n} \bar{s}_h.
    \label{eq_system_losses}
\end{align}
This variable is a global quantity that continues to be widely used in power systems to assess overall system efficiency.  However, by itself, it does not provide dynamic information on the overall system behavior.  {\color{black}Writing $\bar{s}_{\mathrm{l}} = s_{\mathrm{l}}\, e^{\jmath\theta_{\mathrm{l}}}$ with 
$s_{\mathrm{l}}=|\bar{s}_{\mathrm{l}}|$, $\theta_{\mathrm{l}}=\arg(\bar{s}_{\mathrm{l}})$, 
the  {\color{black}\ac{cf}} of the system losses follows directly from its definition and separates into amplitude and phase components.  The result is the proposed metric:
\begin{align}
\boxed{
    \bar{\eta}_{s_{\mathrm{l}}}
    = \frac{\dot{\bar{s}}_{\mathrm{l}}}{\bar{s}_{\mathrm{l}}}
    = \frac{\dot{s}_{\mathrm{l}}}{s_{\mathrm{l}}} + \jmath\,\dot{\theta}_{\mathrm{l}}
    = \varrho_{s_{\mathrm{l}}} + \jmath\, \omega_{s_{\mathrm{l}}},
} \label{eq_cf_losses}
\end{align}
}
where $\varrho_{s_{\mathrm{l}}}$ denotes the instantaneous relative rate of change of the apparent power loss, useful as a normalized dynamic stress indicator, whereas $\omega_{s_{\mathrm{l}}}$ reflects the balance between real and reactive losses.  Both quantities remain independent of system size.  $\bar{\eta}_{s_\mathrm{l}}$ provides a compact, instantaneous metric of the scale and composition of the dynamic behavior of network losses.

To further analyze the physical meaning and practical application of the proposed metric, one can derive (\ref{eq_cf_losses}) by expanding (\ref{eq_complex_power}) for a quasi-steady state model of the transmission system.  This leads to the well-known power flow equations, as follows:
\begin{align}
    p_h = v_h\, \sum\nolimits_{k}^{n} v_k \left[ G_{hk}\cos{\theta_{hk}}+B_{hk}\sin{\theta_{hk}}\right]\,, \label{eq_power_flow_p}\\
    q_h = v_h\, \sum\nolimits_{k}^{n} v_k \left[ G_{hk}\sin{\theta_{hk}}-B_{hk}\cos{\theta_{hk}}\right]\,, \label{eq_power_flow_q}
\end{align}
where $G_{hk}$ and $B_{hk}$ are the real and imaginary parts of the $(h, k)$ element $\bar{Y}_{hk}$ of the network admittance matrix $\boldsymbol{\bar{{Y}}}$, $v_h$ and $v_k$ are the voltage magnitudes at buses $h$ and $k$, respectively, and $\theta_{hk}$ is the phase difference between $\theta_h$ and $\theta_k$, the voltage phase angles at buses $h$ and $k$, respectively.

Differentiation with respect to time of (\ref{eq_power_flow_p}) and (\ref{eq_power_flow_q}) gives:
\begin{align}
    \dot{p}_h&=\frac{p_h}{v_h}{\dot{v}_h}-q_h\, \dot{\theta}_{h}+\sum\nolimits_{k}^{n}\left(\frac{p_{hk}}{v_k}\dot{v}_{k}+{q_{hk}}\dot{\theta}_{k} \right) ,\\
    \dot{q}_h&=\frac{q_h}{v_h}{\dot{v}_h}+p_h\, \dot{\theta}_{h}+\sum\nolimits_{k}^{n}\left(\frac{q_{hk}}{v_k}\dot{v}_{k}+{p_{hk}}\dot{\theta}_{k} \right) ,    
\end{align}
where {\color{black}
\begin{equation}
    \begin{aligned}
        p_{hk} &=  v_h v_k \left[ G_{hk}\cos{\theta_{hk}}+B_{hk}\sin{\theta_{hk}}\right], \\ q_{hk} &=  v_h v_k \left[ G_{hk}\sin{\theta_{hk}}-B_{hk}\cos{\theta_{hk}}\right].
    \end{aligned}
\end{equation}
Using the notation of the {\color{black}\ac{cf}}, we obtain the expression \cite{Milano2022cmplx}:
\begin{align}
    \dot{\bar{s}}_h=\bar{s}_h \bar{\eta}_{v_h}+\sum\nolimits_{k}^{n}{\bar{s}_{hk}\bar{\eta}_{v_k}^{*}} ,
    \label{eq_cmplx_freq_power}
\end{align}
where $\bar{\eta}_{v_h}$ and $\bar{\eta}_{v_k}$ are the \acp{cf} of the voltage at buses $h$ and $k$, and $\bar{s}_{hk} = p_{hk}+\jmath \, q_{hk}$.  The first term of the right-hand side of equation (\ref{eq_cmplx_freq_power}) refers to the total variation of the complex power due to local changes in the voltage, which is related to the devices connected to the bus itself; the second term captures the effect of network topology on power variations at bus $h$.  

The sum of the currents incident at bus $h$ is given by: 
\begin{equation}
    \bar{\imath}_h=\sum\nolimits_k^n \bar{Y}_{hk}\, \bar{v}_k .
\end{equation}
Differentiation with respect to time in terms of \acp{cf} leads to:
\begin{equation}
    \begin{aligned}
    \bar{\imath}_h \bar{\eta}_{\imath_h}
    &=\sum\nolimits_k^n \bar{Y}_{hk}\, \bar{v}_k \, \bar{\eta}_{v_k} .  
    \end{aligned}
\end{equation}
Then, conjugation and multiplication by $\bar{v}_h$ give:
\begin{equation}
   \begin{aligned}
    \bar{v}_h{\bar{\imath}}_h^*  \bar{\eta}_{\imath_h}^* 
    &=\bar{v}_{h}\sum\nolimits_k^n \bar{Y}_{hk}^* \, \bar{v}_k^* \, \bar{\eta}_{v_k}^*,  
    \end{aligned}
\end{equation}
or, equivalently, in terms of complex powers:
\begin{equation}
    \bar{s}_h\bar{\eta}_{\imath_h}^*=\sum\nolimits_k^n \bar{s}_{hk} \, \bar{\eta}_{v_k}^*, \label{eq_weighted_current_cf}
\end{equation}
where $\bar{\imath}_h$ and $\bar{\eta}_{\imath_h}$ are the net current and its \ac{cf} at bus $h$.  To obtain the proposed metric in terms of the \acp{cf} of voltages and currents, we substitute \eqref{eq_weighted_current_cf} into \eqref{eq_cmplx_freq_power}:
\begin{equation}
    \dot{\bar{s}}_h = \bar{s}_h \bar{\eta}_{v_h} + \bar{s}_h \bar{\eta}_{\imath_h}^*
    = \bar{s}_h \left( \bar{\eta}_{v_h} + \bar{\eta}_{\imath_h}^* \right).
    \label{eq_dot_sh_step1}
\end{equation}

Summing \eqref{eq_dot_sh_step1} over $h$ and using \eqref{eq_system_losses} gives the time derivative of the total complex losses:
\begin{equation}
    \dot{\bar{s}}_{\mathrm{l}} = \sum_{h}^{n} \dot{\bar{s}}_h
    = \sum_{h}^{n} \bar{s}_h \left( \bar{\eta}_{v_h} + \bar{\eta}_{\imath_h}^* \right).
    \label{eq_dot_sl_step2}
\end{equation}

Next, we normalize $\dot{\bar{s}}_{\mathrm{l}}$ by dividing \eqref{eq_dot_sl_step2} by $\bar{s}_{\mathrm{l}} = \sum_{h}^{n} \bar{s}_h$.  Recalling the definition of the \ac{cf} in \eqref{eq_cf_losses}, the resulting expression yields:
\begin{align}
    \bar{\eta}_{s_{\mathrm{l}}}
    = \frac{\dot{\bar{s}}_{\mathrm{l}}}{\bar{s}_{\mathrm{l}}}
    = \frac{\sum_{h}^{n} \bar{s}_h \bar{\eta}_{v_h}}{\sum_{h}^{n} \bar{s}_h}
    + \frac{\sum_{h}^{n} \bar{s}_h \bar{\eta}_{\imath_h}^*}{\sum_{h}^{n} \bar{s}_h},
    \label{eq_cf_losses_v1}
\end{align}
which is the intermediate form of the proposed metric.  The first term is a power-injection-weighted average of the bus voltage \acp{cf}, and the second is a power-injection-weighted average of the conjugate bus current \acp{cf}.

Defining the system weighted \ac{cf} of voltages and net currents as:
\begin{equation}
    \bar{\eta}_{v_{\mathrm{sys}}} = \frac{\sum_{h}^{n} \bar{s}_h \bar{\eta}_{v_h}}{\sum_{h}^{n} \bar{s}_h},
    \qquad
    \bar{\eta}_{\imath_{\mathrm{sys}}}^* = \frac{\sum_{h}^{n} \bar{s}_h \bar{\eta}_{\imath_h}^*}{\sum_{h}^{n} \bar{s}_h},
    \label{eq_sys_cf_def}
\end{equation}
equation \eqref{eq_cf_losses_v1} can be written compactly as
\begin{equation}
    \label{eq_cf_losses_v2}
    \boxed{
    \begin{aligned}
        \bar{\eta}_{s_{\mathrm{l}}} 
        &= \bar{\eta}_{v_\mathrm{sys}}+\bar{\eta}_{\imath_\mathrm{sys}}^* \\
        & = \varrho_{v_{\mathrm{sys}}}+\varrho_{\imath_{\mathrm{sys}}} + \jmath\,(\omega_{v_{\mathrm{sys}}}-\omega_{\imath_{\mathrm{sys}}}) ,
    \end{aligned}
    }
\end{equation}
where $\bar{\eta}_{v_{\mathrm{sys}}}$ and $\bar{\eta}_{\imath_{\mathrm{sys}}}$ refer to the system-weighted {\color{black}\ac{cf}} of voltages and net currents across the network buses, for which $\varrho_{v_{\mathrm{sys}}}$,  $\varrho_{\imath_{\mathrm{sys}}}$, $\omega_{v_{\mathrm{sys}}}$ and $\omega_{\imath_{\mathrm{sys}}}$ represent their real and imaginary parts, respectively.  

In \eqref{eq_cf_losses_v2}, $\bar{\eta}_{v_{\mathrm{sys}}}$ is a local device-driven metric.  In particular, the term $\omega_{v_{\mathrm{sys}}}$ can be interpreted as a generalization of the conventional frequency of the \ac{coi}, with the property of being independent of device technology and, thus, it does not require information on the inertia constants of the machines.  The term $\varrho_{v_{\mathrm{sys}}}$ reflects voltage variations at the most `influential' buses, capturing the aggregate effect of voltage control.  On the other hand, $\bar{\eta}_{\imath_{\mathrm{sys}}}$ is a network-driven metric and captures how current magnitudes and phases through transmission lines are responding to the voltage dynamics, which reflects the propagation of disturbances across the grid.  
Finally, note that, in steady-state, the metric $\bar{\eta}_{s_{\mathrm{l}}}$ tends to zero, which also reflects the inherent local synchronization of devices with the grid \cite{localsync} through a coherent network response.

It is relevant to note that this metric is valid also for dc grids.  However, for these types of networks, the imaginary part of (\ref{eq_cf_losses_v2}) is zero as $q_h=0$, and $\omega_{v_h}=\omega_{\imath_h}=0$ for all $h$.

\section{Case Study}


This section presents the dynamic behavior of the proposed metrics in two test systems.  The first is a modified IEEE 39-bus system, used to study the effect of network $X/R$ ratio and of limited voltage/frequency support from \acp{ibr} under a single load outage.  The second is a detailed all-island model of the Irish transmission system, used to assess the metric under repeated, uncoordinated data-center load variations in a realistic large-scale network.  All simulations were performed using the Dome software tool \cite{dome}.

\subsection{IEEE 39-bus System (Contingency/Abnormal Operation)}
\label{sec:case_39bus}

The original 10 \acp{sm} of the IEEE 39-bus system have been replaced with a combination of {\color{black}80}\% \ac{gfl} and {\color{black}20}\% \ac{gfm} \ac{vsm} \acp{ibr} with droop frequency and voltage control.  {\color{black}Three cases are analyzed: Case 1 (base) with $R/X=0.1$ and \ac{gfl} supporting voltage and frequency control; Case 2 increases the $R/X$ ratio to 1, while maintaining the magnitude of each line impedance, to assess voltage-frequency coupling in the grid; Case 3 keeps $R/X=0.1$ but \ac{gfl} devices provide no frequency or voltage support.} 

Figure~\ref{fig_39_bus} shows the magnitude of the apparent power losses, frequency of the \ac{coi}, the real and imaginary parts of the local voltage and network current components, and the \ac{cf} of the losses, for a load outage at bus 8 {\color{black}across all cases.}

\begin{figure*}[t]
    \centering
    \includegraphics[width=0.925\linewidth]{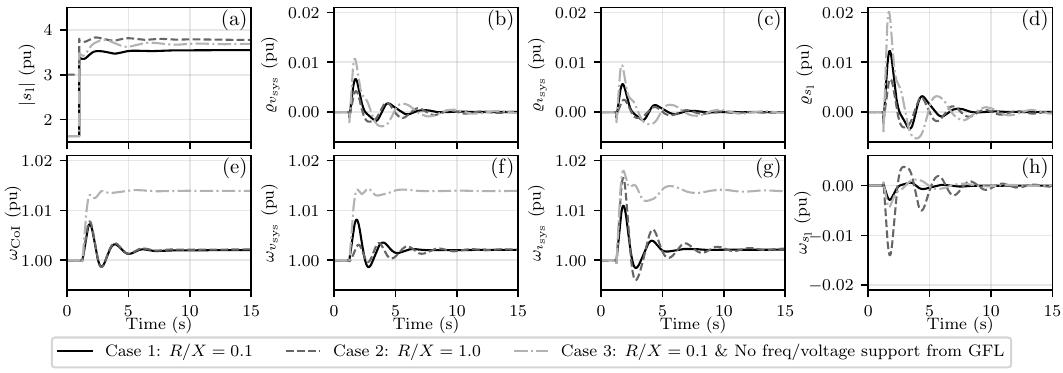}
    \caption{IEEE 39-bus system: load outage at bus~8 for Case 1: $R/X=0.1$, Case 2: $R/X=1$, and Case 3: $R/X=0.1$ and no frequency/voltage support from \acp{gfl}.  Panels (a,e): system loss magnitude \& \ac{coi} frequency; (b,f): real/imaginary parts of the \ac{cf} of local components; (c,g): real/imaginary parts of the \ac{cf} of network components; (d,h): real/imaginary parts of the \ac{cf} of losses.}
    \label{fig_39_bus}
\end{figure*}

{\color{black}In conventional systems ($R/X \approx 0.1$), the frequency of the \ac{coi} $\omega_{\mathrm{CoI}}$, as defined in \cite{FERNANDEZ2019} for systems with \ac{gfm}-\ac{vsm} converters, coincides with the steady-state frequency and is comparable to $\omega_{v_{\mathrm{sys}}}$ during transients.  However, as the $R/X$ ratio increases, voltage control increasingly distorts frequency indices as the influence of coupled voltage–frequency terms with complex power is amplified, while not affecting the conventional frequency of the \ac{coi} nor the dynamic behavior of system losses, which are substantially identical in both scenarios except for an offset due to different line resistances.

Comparing Cases 1 and 2 with, respectively, $R/X = 0.1$ and $1.0$, the \ac{rocof} at 500 ms of $\omega_{\mathrm{CoI}}$ remains practically unchanged at around 0.009 pu/s, whereas $\omega_{v_{\mathrm{sys}}}$ decreases significantly from 0.009 to 0.001 pu/s, indicating a lower \ac{rocof} and a higher frequency nadir due to the impact of voltage support from generators in the more coupled case.  Additionally, $\omega_{\imath_{\mathrm{sys}}}$ increases from 0.014 to 0.023 pu/s, making the network-driven component dominant in $\omega_{s_{\mathrm{l}}}$ and suggesting, according to \cite{localsync}, that synchronism becomes more difficult to achieve.  In these cases, the real-part metrics $\varrho_{v_{\mathrm{sys}}}$, $\varrho_{\imath_{\mathrm{sys}}}$ and $\varrho_{s_{\mathrm{l}}}$ exhibit a slight reduction in first-swing amplitude with increasing $R/X$ due to normalization by increased active and apparent losses, all returning to zero toward steady state.

Without voltage/frequency support from \acp{gfl} inverters (Case~3, $R/X = 0.1$), $\omega_{\mathrm{CoI}}$ leads to a lower \ac{rocof} at 500 ms than $\omega_{v_{\mathrm{sys}}}$ (0.0148 pu/s versus 0.0162 pu/s).  Noticeably, $\varrho_{v_{\mathrm{sys}}}$ and $\varrho_{\imath_{\mathrm{sys}}}$ become much larger than in the previous cases, showcasing the importance of adequate voltage/frequency support from \acp{gfl}.  Table~\ref{tab:rocof_zenith_39bus} summarizes the \ac{rocof} at 500~ms and the frequency zenith values of the \ac{coi} frequency $\omega_{\mathrm{CoI}}$ and of the proposed metrics $\omega_{v_{\mathrm{sys}}}$ and $\omega_{\imath_{\mathrm{sys}}}$ for the three cases.

\begin{table}[t]
\caption{IEEE 39-bus System: Summary of \ac{rocof} at 500~ms and Zenith.}
\centering
\footnotesize
\setlength{\tabcolsep}{4pt}
\begin{tabular}{@{}l c c r r r r r r}
\toprule
 & & \ac{gfl}  & \multicolumn{3}{c}{\ac{rocof} at 500~ms (pu/s)} & \multicolumn{3}{c}{Zenith (pu)}\\
\cmidrule(lr){4-6} \cmidrule(lr){7-9}
Case & $R/X$ & support & $\dot{\omega}_{\mathrm{CoI}}$ & $\dot{\omega}_{v_{\mathrm{sys}}}$ & $\dot{\omega}_{\imath_{\mathrm{sys}}}$ & $\omega_{\mathrm{CoI}}$ & $\omega_{v_{\mathrm{sys}}}$ & $\omega_{\imath_{\mathrm{sys}}}$ \\
\midrule
1 & $0.1$ & $\checkmark$ & $ 0.009$ & $ 0.009$ & $ 0.014$ & $1.008$ & $1.008$ & $1.011$ \\
2 & $1.0$ & $\checkmark$ & $ 0.009$ & $ 0.001$ & $ 0.023$ & $1.008$ & $1.003$ & $1.016$ \\
3 & $0.1$ & $\times$     & $ 0.015$ & $ 0.016$ & $ 0.023$ & $1.013$ & $1.014$ & $1.018$ \\
\bottomrule
\end{tabular}
\label{tab:rocof_zenith_39bus}
\end{table}

The components $\varrho_{\rm s_l}$ and $\omega_{\rm s_l}$ reflect the system's post-contingency dynamic synchronization capability---see Fig.~\ref{fig_39_bus}(d) and (h).  For $\varrho_{\rm s_l}$, Case 3 is the most critical scenario, highlighting GFL voltage support as a crucial system resource. Conversely, for $\omega_{\rm s_l}$, Case 2 is the most critical, indicating that a weak grid primarily degrades angular frequency synchronization due to heightened sensitivity (driven by $R$) between reactive power and angular frequency.

\subsection{All-island Irish Transmission System (Normal Operation)}
\label{sec:case_ireland}
To evaluate the proposed metric in a realistic large-scale network with fast and repeated load perturbations, the second case study is conducted on the all-island model of the Irish transmission system \cite{ponce2026}.  The case comprises approximately 1976 buses and 1776 branches.  Synchronous generation is represented by a 6th-order machine model including associated excitation and governor controls, while \acp{ibr} are modeled with WECC-type REGC/REECA plant controls \cite{epriwecc}.  The island is connected to Great Britain through the EWIC, Greenlink and Moyle HVDC interconnectors.  


In the simulated scenario, the system dynamics are driven by data-center demand.  Ten aggregate data-center loads are connected in the Dublin area at the low-voltage side of existing transformers.  Each load is modeled as described in \cite{DCmodel} and comprises a CPU, a pulsed GPU, a ZIP load, and a cooling induction motor with zero mechanical torque.  The sites are distributed in four stations: three at Clonee ($20\,\mathrm{kV}$), two at Belcamp ($10.5\,\mathrm{kV}$), three at Corduff ($10.5\,\mathrm{kV}$), and two at Carrickmines ($10.5\,\mathrm{kV}$).  Total peak data-center demand is approximately $800\,\mathrm{MW}$, with individual ratings in the range $60$ to $100\,\mathrm{MW}$.  Each data center comprises a CPU/base load and a GPU share of about $20\%$ of its rating, plus a small component of about $1\%$.  GPU demand is modulated by periodic pulses, with first-order lags $T_{\mathrm{cpu}}=1\,\mathrm{s}$ and $T_{\mathrm{gpu}}=0.1\,\mathrm{s}$, cycle periods drawn randomly in the range $15$ to $40\,\mathrm{s}$, and ON duty cycles in the range $0.7$ to $0.9$.  Pulse phases are randomized independently across data centers. 

Figure~\ref{fig:datacenters} shows the stacked data-center consumption of active power.  Although parameters are randomized for their operation cycles, variations up to 100 MW can be observed when the cycles coincide (e.g.  at 350 s). 

\begin{figure}[t]
    \centering
    \includegraphics[width=0.93\linewidth]{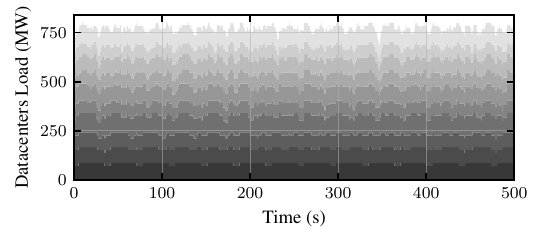}
    \caption{Stacked active-power demand of the ten data centers.}
    \label{fig:datacenters}
\end{figure}

Figure~\ref{fig:ireland}(a) shows the frequency in Hz of the local and network components of the proposed metric, together with the \ac{coi} frequency. The two components almost overlap, with only small differences mainly due to fast voltage changes and dynamics driven by non-synchronous devices. These dynamics are captured by both $\omega_{v_{\mathrm{sys}}}$ and $\omega_{\imath_{\mathrm{sys}}}$ but not by the \ac{coi} frequency, as also observed in Section~\ref{sec:case_39bus}. Figure~\ref{fig:ireland}(b) shows the complex frequency of losses $\bar{\eta}_{s_{\mathrm{l}}}=\varrho_{s_{\mathrm{l}}}+\jmath\omega_{s_{\mathrm{l}}}$ in pu. The real part $\varrho_{s_{\mathrm{l}}}$ exhibits pronounced high-frequency content associated with fast voltage and non-synchronous dynamics, whereas the imaginary part $\omega_{s_{\mathrm{l}}}$ captures some of the fast voltage-driven changes but mainly follows the slower network dynamics. Both $\varrho_{s_{\mathrm{l}}}$ and $\omega_{s_{\mathrm{l}}}$ vary around zero, indicating that the system continuously seeks synchronism.

Fast load variations are translated almost directly into voltage variations at the data center connection buses, as shown in Fig.~\ref{fig:ireland}(c).  Figure~\ref{fig:ireland}(d) shows the autocorrelation of the voltages at representative buses in the Irish grid, including the integral of the $\varrho$ part of the metrics.  The integral is used so that results are comparable with the voltage dynamics.  $\varrho_{v_{\mathrm{sys}}}$ captures the voltage dynamics, and its integrated autocorrelation indicates that it can be a good measure for assessing voltage dynamics within a unified network metric, as it represents the overall behavior of the voltages throughout the system.  $\varrho_{\imath_{\mathrm{sys}}}$ provides additional information about voltage propagation; its autocorrelation remains almost always higher than the others, mainly because the network component smooths the fast changes observed in the local component.  

\begin{figure*}[t]
    \centering
    \includegraphics[width=0.9\linewidth]{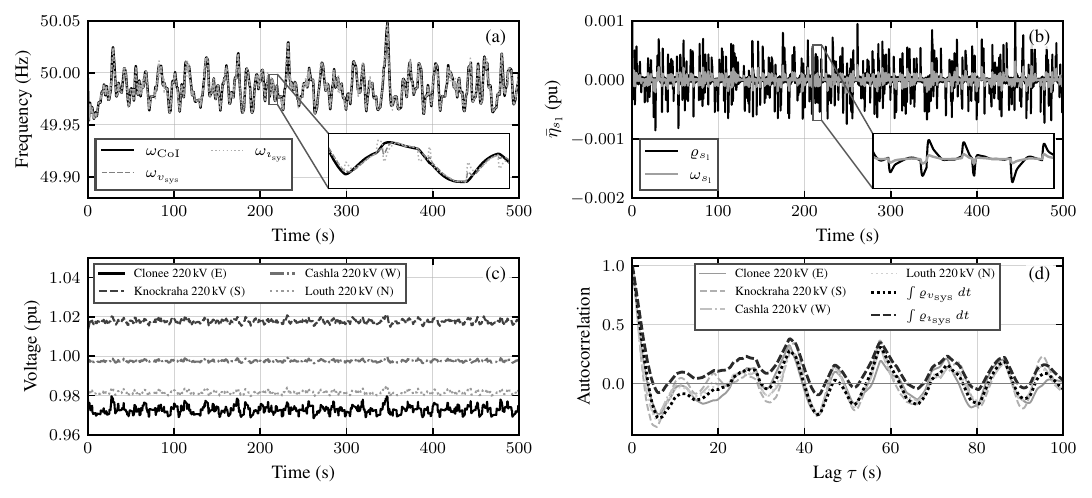}
    \caption{Dynamic response of the all-island Irish system with ten Dublin \acp{dc} over $500\,\mathrm{s}$.
(a)~System frequencies: voltage- and current-based $\omega_{v_{\mathrm{sys}}}$, $\omega_{\imath_{\mathrm{sys}}}$ and \ac{coi} frequency $\omega_{\mathrm{CoI}}$. (b)~Complex frequency of losses $\bar{\eta}_{s_{\mathrm{l}}}=\varrho_{s_{\mathrm{l}}}+\jmath\omega_{s_{\mathrm{l}}}$. (c)~Bus voltages at four buses: Clonee (east), Knockraha (south), Cashla (west), and Louth (north). (d)~Autocorrelation of the same bus voltages along with the autocorrelations of $\int\varrho_{v_{\mathrm{sys}}} dt$ and $\int\varrho_{\imath_{\mathrm{sys}}} dt$.}
    \label{fig:ireland}
\end{figure*}

\subsection{Remarks on Simulation Results}

A key strength of the proposed metric lies in its exact theoretical decomposition into device-driven and network-driven components, enabling a comprehensive assessment of local interactions versus grid-wide perturbation propagation.  Simulation results demonstrate that, in traditional synchronous-machine-dominated systems, the imaginary part of the device-driven component closely aligns with the conventional frequency of the center of inertia (\ac{coi}).  However, in modern grids exhibiting strong voltage-frequency coupling, the proposed metric extracts relevant dynamic information that traditional \ac{coi}-based metrics fail to capture.  This capability is validated on both a modified IEEE 39-bus test system and the realistic all-island Irish transmission grid, where the metric adequately characterizes system dynamics under high \ac{ibr} penetration and fast data-center load variations.

Finally, a qualitative summary of the features of $\omega_{\mathrm{CoI}}$, \ac{cf} and the proposed metric is provided in Table~\ref{tab:metrics}.

\begin{table}[htbp]
\caption{Comparison of metrics by attributes.}
\centering
\footnotesize
\setlength{\tabcolsep}{3pt}
\newlength{\colwidth}
\setlength{\colwidth}{\dimexpr 84mm/4\relax}
\begin{tabular}{
    p{1.15\colwidth}
    >{\raggedright\arraybackslash}p{\colwidth}
    >{\raggedright\arraybackslash}p{0.8\colwidth}
    >{\raggedright\arraybackslash}p{0.9\colwidth}
}
\toprule
{Attribute} & ${\omega_{\mathrm{CoI}}}$ & {CF} & {Proposed} \\
\midrule
Local/System-wide & System-wide & Local & System-wide \\ \\
Technology Agnostic & No & Yes & Yes \\ \\
Dynamics considered & Avg.~frequency deviations of \ac{sm} & Local voltage dynamics & Power flows \& synchronization \\
\bottomrule
\end{tabular}
\label{tab:metrics}
\end{table}

\section{Conclusions}
This paper proposes a novel system-wide dynamic metric to evaluate the stability and performance of modern power systems.  By accounting for all network components that influence power flows, the proposed metric is inherently technology-agnostic and well-suited for grids with high penetration levels of \acp{ibr}.  Formulated as the complex frequency of total system losses, the metric captures both active and reactive power balances alongside their associated dynamics through its real part ($\varrho_{s_{\mathrm{l}}}$) and imaginary part ($\omega_{s_{\mathrm{l}}}$), while the condition $\bar{\eta}_{s_{\mathrm{l}}}=0$ serves as an indicator of full system synchronization.

Future research will focus on extending this unified framework along three main directions: adapting the dynamic formulation to multi-terminal dc and hybrid ac-dc power networks, developing robust estimation techniques using sparse phasor measurement unit (PMU) data to enable online implementation under limited measurement observability, and engaging with system operators to establish standard guidelines for integrating the metric into dynamic security assessment protocols and frequency quality monitoring tools.



\end{document}